\documentclass[aps,prd,twocolumn,floatfix,noshowpacs,tightenlines,noshowkeys,superscriptaddress,amsmath,amssymb,nofootinbib]{revtex4}
\usepackage{amssymb,amsbsy,epsfig,color,graphicx}
\usepackage{color}
\usepackage{[longtable}
\usepackage{subfigure}
\usepackage{array}
\usepackage{dcolumn}   
\usepackage{cellspace}
\usepackage{mathtools}
\usepackage{amstext}
\usepackage{amssymb}
\usepackage{stmaryrd}
\usepackage{stackrel}
\usepackage{graphicx}
\usepackage[utf8]{inputenc}
\usepackage{blindtext}
\usepackage{float}
\restylefloat{table}
\usepackage{booktabs}
\usepackage{enumitem}
\usepackage{lipsum}
\usepackage{xcolor}

\usepackage{etoolbox} 
\usepackage{lipsum} 
\usepackage[capitalize]{cleveref}
\usepackage{multirow}
\usepackage[caption=false]{subfig}
\renewcommand\[{\begin{equation}}
\renewcommand\]{\end{equation}}

\newcommand{\ba}{\begin{eqnarray}}
\newcommand{\ea}{\end{eqnarray}}

\makeatletter

\appto{\appendix}{%
\@ifstar{\def\theequation@prefix{A.}}%
{}%
}
\makeatother

\begin{document}

\title{Testing fundamental physics with photon frequency shift}


\author{Luca Buoninfante}
\affiliation{Dipartimento di Fisica "E.R. Caianiello", Universit\`a di Salerno, I-84084 Fisciano (SA), Italy}
\affiliation{INFN - Sezione di Napoli, Gruppo collegato di Salerno, I-84084 Fisciano (SA), Italy}
\affiliation{Van Swinderen Institute, University of Groningen, 9747 AG, Groningen, The Netherlands}


\author{Gaetano Lambiase}
\affiliation{Dipartimento di Fisica "E.R. Caianiello", Universit\`a di Salerno, I-84084 Fisciano (SA), Italy}
\affiliation{INFN - Sezione di Napoli, Gruppo collegato di Salerno, I-84084 Fisciano (SA), Italy}

\author{Antonio Stabile}
\affiliation{Dipartimento di Fisica "E.R. Caianiello", Universit\`a di Salerno, I-84084 Fisciano (SA), Italy}
\affiliation{INFN - Sezione di Napoli, Gruppo collegato di Salerno, I-84084 Fisciano (SA), Italy}


\begin{abstract}

We propose a high precision satellite experiment to further test Einstein's General Relativity and constrain extended theories of gravity. We consider the frequency shift of a photon radially exchanged between two observers located on Earth and on a satellite in circular orbit in the equatorial plane. In General Relativity there exists a peculiar satellite-distance at which the static contribution to the frequency shift vanishes since the effects induced by pure gravity and special relativity compensate, while it can be non-zero in modified gravities, like in models with screening mechanisms. As an experimental device placed on the satellite we choose a system of hydrogen atoms which can exhibit the $1$s spin-flip transition from the singlet (unaligned proton-electron spins) to the triplet (aligned proton-electron spins) state induced by the absorption of photons at $21.1$cm. The observation of an excited state would indicate that the frequency of the emitted and absorbed photon remains unchanged according to General Relativity. On the contrary, a non-zero frequency shift, as predicted in extended theories of gravity, would prevent the spin-flip transition and the hydrogen atoms from jumping into the excited state.
Such a detection would signify a smoking-gun signature of new physics beyond special and general relativity.
\end{abstract}

\maketitle


\section{Introduction and physical setup} 

Einstein's general relativity (GR) has been the best theory of the gravitational interaction so far, indeed its predictions have been tested to very high precision \cite{-C.-M.,Turyshev:2008ur}. Despite its great success there are still open questions which make the theory incomplete. In fact, at short distances and small time scales Einstein's GR predicts black hole and cosmological singularities, which signal the presence of spacetime points at which predictability is lost. Moreover, cosmological and astrophysical observations \cite{riess} show inconsistencies with the theoretical predictions, and new physics in the matter sector has been invoked in order to explain the experiments, i.e. dark energy and dark matter. Recently, it was shown that to match the experimental data, and in particular to solve the so-called cosmological-constant problem, an alternative approach consists in extending Einstein's GR, namely in modifying the nature of the gravitational interaction (or, in other words, the spacetime geometry). In such models, gravity shows a different behaviour either below (ultraviolet modification) or above (infrared modification) a certain length scale, while still keeping all known and well tested properties of GR. One may consider, for example, generalization of Einstein's GR where the Lagrangian is not simply a linear function of the Ricci scalar, $\mathcal{L}\sim R,$ but it can be a more general function of the higher order curvature invariants, $\mathcal{L}\sim f(R,R_{\mu\nu}R^{\mu\nu},\dots)$ \cite{starobinski,Boehmer:2007kx,waterhouse,odi,capoz1}.
%
In relation to the cosmological constant problem, over the past decade a series of
theories has been proposed in which deviations from GR occur only in the {\it ultraweak-field} regime \cite{joyce} through
screening effects.
The latter are realized by introducing an additional degree of freedom, typically represented by a scalar field, that obeys a non-linear equation driven by the matter density, hence coupled to the environment.
Screening mechanisms allow to circumvent Solar system and laboratory tests by dynamically suppressing deviations from GR.
More precisely, the effects of the scalar field is hidden, in high-density regions, by the coupling of the field with matter
while they are unsuppressed and significant on cosmological scales, namely in low-density regions. Well-studied screening mechanisms are the chameleon \cite{veltman}, symmetron \cite{symmetron}, and Vainshtein \cite{Vainshtein}.
New tests of the gravitational interaction may therefore provide an answer to these fundamental questions.
%

In this Letter, we propose a {\it novel} experiment to further test and constrain the real nature of gravity. We consider a physical setup in which a photon is exchanged between two observers: the observer $A$ is sitting on Earth of radius $r_A$ and angular velocity $\omega_A,$ and sends a photon to the observer $B$ located on a satellite in circular orbit around the Earth at a distance $r_B\,.$ In general, in a curved background and due to the motion of the satellite, the frequency of the photon measured by $B$ will differ from the one measured by $A\,.$ 

In Einstein's GR there exists a special configuration, $r_B=\frac{3}{2}r_A,$ at which the (static) gravitational contribution and the one due to the motion of the satellite compensate and give a vanishing frequency shift, meaning that the two observers clocks tick at same rate, up to corrections of order $\mathcal{O}(\omega_A^2r^2_A).$ However, in extended models of gravity such a compensation may, in general, not occur. Indeed, for any gravitational source, we can determine an observational window in which any kind of non-zero detectable effect would imply the existence of new physics beyond either GR or special relativity. 

As an experimental device we propose a system of hydrogen atoms which can exhibit spin-flip transitions in the $1$s level when absorbing a photon of frequency $1420$MHz ($21.1$cm). By preparing the experiment with this initial photon, in Einstein's GR we would expect an excitation, i.e. the spin-flip transition from the singlet state to the triplet state, while some extended theories of gravity would predict a non-zero frequency shift which would prevent the atoms from jumping into the excited state. A satellite experiment designed for this peculiar physical configuration might offer an extremely suitable and unique scenario to probe new physics.

\section{Theoretical framework}\label{uncert-sec}

We consider a spherical slowly rotating gravitational source of mass $m$ and angular momentum $J,$ whose surrounding spacetime geometry is well described by the linearized metric\footnote{We adopt the mostly positive convention for the metric signature, $\eta={\rm diag}(-1,+1,+1,+1),$ and mainly use the Natural Units, $\hbar=1=c.$}
\begin{equation}
\begin{array}{rl}
ds^2=& \displaystyle -(1+2\Phi(r))dt^2+2\chi(r)\,{\rm sin}^2\theta\,d\varphi\,dt\\[3mm] &\displaystyle\,\,\,\,\,\,\,\,\,\,\,\,+(1-2\Psi(r))(dr^2+r^2d\Omega^2)\,.\label{polar-metric}
\end{array}
\end{equation}
In Einstein's GR, the metric in Eq.\eqref{polar-metric} corresponds to the Lense-Thirring metric \cite{Lense}, indeed we have $\Phi=\Psi=-Gm/r$ and $\chi=-2GJ/r;$ while in extended models of gravity the three metric potentials may have a very different form as we will see below when discussing some applications.

Let us denote $\nu_X$ the frequency of the photon measured by the observer $X$ with proper time $\tau_X,$ where $X=A,B.$ We define the frequency shift of a photon emitted by the observer $A$ and received by the observer $B$ through the formula \cite{Kopeikin:1999ev,Blanchet:2000pn,Kohlrus:2015szb}:
\begin{equation}
\delta\equiv 1-\sqrt{\frac{\nu_B}{\nu_A}},\quad \quad \frac{\nu_B}{\nu_A}=\frac{\left.\left[k_{\mu}u^{\mu}_B\right]\right|_{r=r_B}}{\left.\left[k_{\mu}u^{\mu}_A\right]\right|_{r=r_A}}\,,\label{freq-shift}
\end{equation}
where $k^{\mu}= (\dot{t}_{\gamma},\dot{r}_{\gamma},\dot{\theta}_{\gamma},\dot{\varphi}_{\gamma})$ is the photon four-velocity, while $u^{\mu}_X= (\dot{t}_{X},\dot{r}_X,\dot{\theta}_X,\dot{\varphi}_X)$ are the four-velocities of the two observers $A$ and $B;$ the dots stand for derivatives with respect to the proper time. We study the photon exchange in the equatorial plane, i.e. $\theta=\pi/2$ and $\dot{\theta}_{\gamma}=\dot{\theta}_X=0,$ and assume the orbits of the two observers to be circular, i.e. $\dot{r}_X=0;$ moreover, we choose a configuration in which the photon is sent radially from $A$ to $B$, i.e. $\dot{\varphi}_{\gamma}=0\,.$ Therefore, we can write the two measured frequencies in Eq.\eqref{freq-shift} as 
\begin{equation}
k_{\mu}u^{\mu}_X=\dot{t}_{\gamma}\left(g_{tt}\dot{t}_X+g_{t\varphi}\dot{\varphi}_X\right)\,.\label{scal-prod}
\end{equation}
%
%
%
%
%

The photon is sent radially, thus its four-velocity has only two non-vanishing components
\begin{equation}  k^{\mu}=\left((1-2\Phi)E_{\gamma},\dot{r}_{\gamma},0,0\right)\,,\label{eneg-angul-prop}
\end{equation}
where $E_{\gamma}=\displaystyle (1+2\Phi)\dot{t}_{\gamma}-\chi \dot{\varphi}_{\gamma}$ is the conserved energy of the photon along its geodesic. The four-velocities of the two observers $A$ and $B$ are \cite{chandra}
\begin{equation}
u_X^{\mu}= \displaystyle \left.\frac{(1,0,0,\omega_X)}{\sqrt{(1+2\Phi)-(1-2\Psi)r^2\omega_X^2-2\chi \omega_X}}\right|_{r=r_X}\,.
\label{4-vel-B}
\end{equation}
The quantity $\omega_A=\dot{\varphi}_A/\dot{t}_A$ is the angular velocity of the observer $A$ which is not following a geodesic, indeed it corresponds to source's equatorial angular velocity; while $\omega_B=\dot{\varphi}_B/\dot{t}_B$ is the angular velocity of the observer $B$ on the satellite which follows a geodesic and it can be expressed in terms of the Christoffel symbols by explicitly solving the geodesic equation for the component $\mu=r\,.$ Note that in both cases the normalization factor has been fixed by imposing $u_X^{\mu}(u_X)_{\mu}=-1,$ with $X=A,B\,.$

The frequency shift in Eq.\eqref{freq-shift}, up to linear order in the metric potentials and in the slow angular velocity regime, reads:
\begin{equation}
\delta=\delta_{\rm stat}+\delta_{\rm rot}\,,
\label{shift-degree}
\end{equation}
where
\begin{equation}
\begin{array}{rl}
\delta_{\rm stat}\equiv& \displaystyle  -\frac{1}{2}\left[\Phi(r_A)-\Phi(r_B)+\frac{r_B\Phi^{\prime}(r_B)}{2}\right],\\[3.7mm]
\delta_{\rm rot}\equiv& \displaystyle  \frac{r_A^2\omega_A^2}{4}\left[1-\frac{3}{2}\Phi(r_A)-\frac{\Phi(r_B)}{2}\right.\\
&\displaystyle \,\,\,\,\,\,\,\,\,\,\,\,\,\,\,\,\,\,\,\,\,\,\,\,\,\,\,\,\left.+\frac{r_B\Phi^{\prime}(r_B)}{4}-2\Psi(r_A)\right]\,,
\end{array}
\label{static-rot}
\end{equation}
are the static and rotational contributions, respectively.

\begin{figure}[t!]
	\includegraphics[scale=0.325]{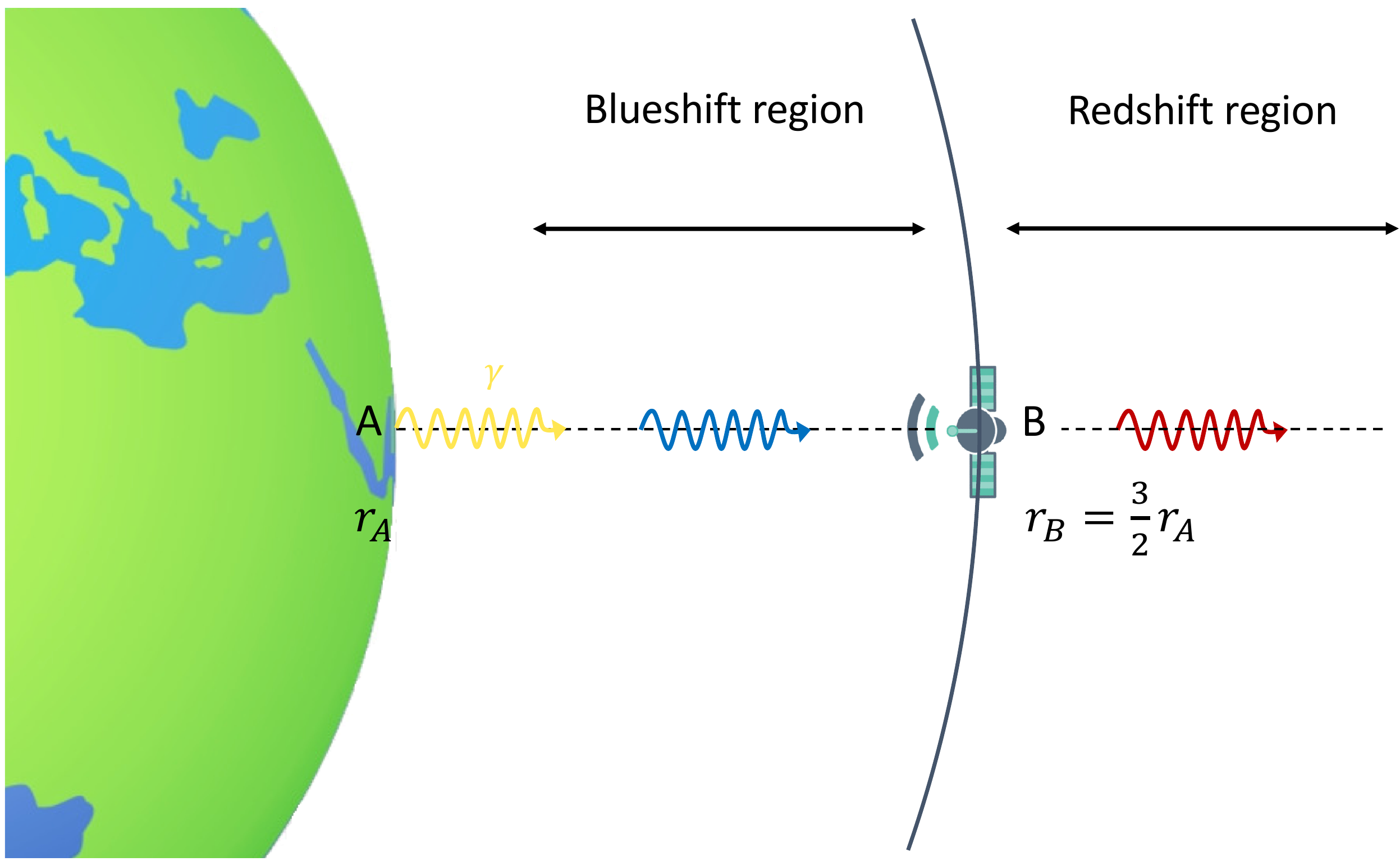}
	\caption{We have illustrated what happens to a photon sent by an observer $A$ on Earth and received by an observer $B$ on a satellite in circular orbit. The formula for the frequency shift of the photon is given by Eq.\eqref{static-rot} and tells us that in GR one has the pure static gravitational contribution, $\Delta\Phi=\Phi(r_A)-\Phi(r_B),$ plus a piece related to the geodesic motion of the satellite, $r_B\Phi^{\prime}(r_B)/2\simeq Gm/(2r_B),$ which can be interpreted as a special relativistic contribution. For $r_B<\frac{3}{2}r_A$ the photon is seen blue-shifted by the observer $B$ as the contribution due to the satellite motion is dominant over the gravitational gradient $\Delta\Phi;$ for $r=\frac{3}{2}r_A$ the two effects compensate each other; for $r_B>\frac{3}{2}r_A$ the term $\Delta\Phi$ dominates over the special relativistic one and the photon is seen red-shifted by the observer $B\,.$ }
	\label{fig12-2}
\end{figure}

We can immediately notice that in GR there exists a peculiar configuration, namely $r_B=\frac{3}{2}r_A,$ for which there is no static contribution to the frequency shift. Indeed, if $\Phi=-Gm/r,$ we have
\begin{equation}
\displaystyle \delta_{\rm stat}^{{\rm GR}}= \displaystyle \frac{Gm}{2}\left(\frac{1}{r_A}-\frac{3}{2}\frac{1}{r_B}\right),
\label{gr-3/2}
\end{equation}
which vanishes for $r_B=\frac{3}{2}r_A\,,$ in agreement with the finding in Ref. \cite{Kohlrus:2015szb}. For the Schwarzschild metric, this peculiar value of the distance corresponds to the location at which the (static) gravitational shift $\Delta\Phi=\Phi(r_A)-\Phi(r_B)$ compensate the one induced by the circular motion of the satellite around the Earth, $r_B\Phi^{\prime}(r_B)/2=Gm/(2r_B)$ \cite{Kohlrus:2015szb}. Note that the latter can be interpreted as a special relativistic contribution induced by the relative motion of the satellite with respect to the observer $A.$ Indeed, by making a simple computation in special relativity of the frequency shift due to the relative motion of the observer $B$ with respect to $A$ would obtain $(\nu_B/\nu_A)_{\rm SR}\simeq 1+Gm/(2r_B)\,.$ For distances $r_B>\frac{3}{2}r_A$ the pure gravitational effect is the dominant one and the photon is seen red-shifted by the observer $B$ on the satellite; while for $r_B<\frac{3}{2}r_A,$ the contribution due to the motion of the satellite dominates so that the observer $B$ sees the photon blue-shifted. See Fig. \ref{fig12-2} for an illustration. Hence, on such a peculiar orbit the first non-vanishing contribution in GR comes from the rotational term $\delta_{\rm rot}\sim \mathcal{O}(\omega_A^2r_A^2).$ Such a property {\it may not} hold in extended theories of gravity. 

%

We emphasize that if an experiment with a satellite in circular orbit at the distance $r_B=\frac{3}{2}r_A$ is performed, then any kind of observed non-vanishing contribution of the order $|\delta|\gtrsim \mathcal{O}(\omega_A^2r_A^2),$ would signify the existence of new physics beyond either GR or special relativity. Therefore, such a novel experimental scenario might be extremely promising in order to improve satellite Solar system tests of GR and to further constrain physics beyond Einstein's theory.
%

\section{Observational windows}

We now want to determine the observational window in which any kind of detectable effect would imply the presence of new physics. To this aim, we need to estimate both the static and the rotational contributions to the frequency shift in Eq.\eqref{static-rot}.

As a first example, let us consider the Earth as the gravitational source on which the observer $A$ is sitting and rotating with angular velocity $\omega_A.$ The radius of Earth is $r_A=6.37\times 10^6$m, its mass $m=5.97\times10^{24}$kg and the angular velocity $\omega_A=7.36\times 10^{-5}$rad/s, therefore the static and rotational contributions are given by $|\delta_{\rm stat}|\simeq \frac{1}{2}Gm/r_A\simeq 3.48\times 10^{-10}$ and $|\delta_{\rm rot}|\simeq \frac{1}{4}\omega_A^2r_A^2\simeq 6.02\times 10^{-13}.$ Hence, the observational window is given by:
\begin{equation}
{\rm Earth:}\quad\quad  6.02\times 10^{-13} \lesssim \left|\delta\right| \lesssim 3.48\times 10^{-10}
\label{window-earth}
\end{equation}
For the Earth the distance at which the static GR contribution to the shift vanishes is $r_B=\frac{3}{2}r_A\simeq 9556.5\,{\rm km}$. Thus, if an experiment with a satellite in circular orbit at such a distance is performed, then any kind of non-vanishing detectable frequency shift falling in the range given by \eqref{window-earth}, would represents a smoking gun signature of new physics.

So far, we have only considered the Earth as an example of gravitational source, but in principle we can go further and apply the same setup to several sources in the Solar system. Experiments which now seem to be impossible might become feasible even in the far future; for instance, it might become possible to realize it with the observer $A$ on the Moon and the observer $B$ on a satellite in circular orbit around it.

Indeed, very interestingly, in the case of the Moon the observational window turns out to be larger as compared to the Earth case due to a smaller radius and a lower angular velocity. The radius of the Moon is $1.74\times 10^{6}\,{\rm m}$, its mass $7.35\times 10^{22}\,{\rm kg}$, while the angular velocity is $\omega_{A}=2.70\times 10^{-6}\,{\rm rad/s}$. We can easily estimate the static and rotational contributions to the shift and obtain the following observational window: $ 1.38\times 10^{-16} \lesssim \left|\delta\right|\lesssim 1.57\times 10^{-11} .$ The peculiar distance for the Moon is $r_B=2605.5$km.

%
\begin{figure*}[t!]
	\includegraphics[scale=0.5]{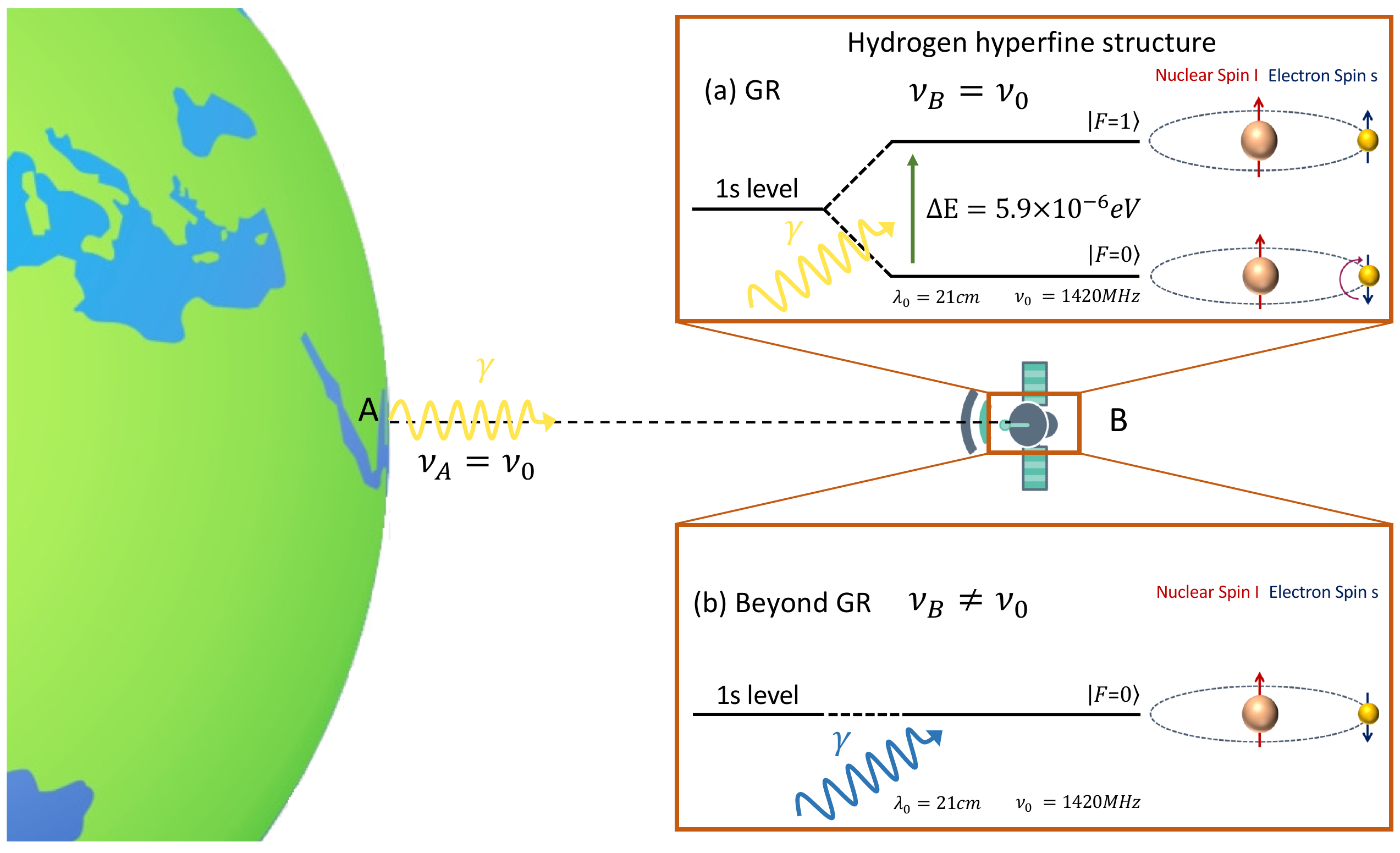}
	\centering
	\caption{Experimental device to measure the photon frequency shift: system of hydrogen atoms which can exhibit a spin-flip transition in the $1$s level. $(a)$ In GR the orbit $r_B=\frac{3}{2}r_A$ corresponds to an Earth-satellite configuration at which the (static) pure gravitational contribution compensates the one related to the relative motion of the satellite with the respect to Earth. The static contribution to the frequency shift is zero (see Eq. (\ref{shift-degree})) up to order $\mathcal{O}(\omega_A^2r^2_A)$. Thus, a photon with frequency $\nu_A=\nu_B=1420$MHz sent by $A$ would arrive in $B$ with the same frequency, $\nu_B=\nu_A,$ and when absorbed by the device would induce an excitation, i.e. a transition from the singlet $|F=0\rangle$ to the triplet $|F=1\rangle$ state. $(b)$ In extended theories of gravity the overall static effect can be non-vanishing since the frequency at the point $B$ can differ from the emitted one at $A,$ indeed it can be either blue-shifted or red-shifted $\nu_B\neq\nu_A\,,$ so that no transition would occur. The latter scenario would signify the existence of new physics beyond GR.}
	\label{fig12}
\end{figure*}
We are working with the linearized form of metric in Eq.\eqref{polar-metric}, which means that we are neglecting any kind of term quadratic and higher in the metric potentials, $\mathcal{O}(\Phi^2)\,.$ However, it is worth emphasizing that such an approximation is well justified and any kind of higher order term in the metric perturbation would lay outside the observational windows we have determined. For example, for the Earth we have $\Phi^2\simeq G^2m^2/r^2_A\simeq 10^{-20},$ which is many order of magnitudes smaller than the corresponding rotational contribution $\omega_{A}^2r_A^2\simeq 10^{-13}.$ The same holds for the other listed gravitational sources, as it can be easily checked.


\section{Experimental device}\label{}

We assume that the observer $B$ on the satellite in circular orbit at $r_B=\frac{3}{2}r_A$ is equipped with a device through which the frequency shift of the photon can be measured.

Let us assume that the excitation frequency of each atom in the system is $\nu_0$ and that the observer $A$ prepares a photon with exactly this frequency, $\nu_A=\nu_0\,.$ Therefore, as long as $\nu_B=\nu_A=\nu_0$ the atomic system will be excited and the observer $B$ would conclude that no frequency shift has been registered. Indeed, this is what happens in GR, since at the distance $r_B=\frac{3}{2}r_A$ the overall frequency shift is zero. However, in extended theories of gravity it can happen that the Newtonian potential is modified such that the gravitational shift does not compensate the special relativistic effect. In this case, the observer $B$ can detect a frequency $\nu_B\neq\nu_0,$ implying that the atoms do not get excited. Therefore, such a device works as a switch and might offer a very favourable experimental scenario to discriminate between different theories of gravity. The relevant quantity is given by $\left|\Delta\nu/\nu_A \right|\simeq 2|\delta|,$ with $\Delta\nu=\nu_B-\nu_A\,$ and, as it happens for $\delta\,,$ it has to lie within the observational window in Eq.\eqref{window-earth} (in the case of Earth) in order for effects beyond GR to be relevant
\begin{equation}
1.20\times 10^{-12} \lesssim \left|\frac{\Delta\nu}{\nu_A}\right| \lesssim 6.96\times 10^{-10} \,.\label{freq-window}
\end{equation}

A possibility along this line is offered by the hydrogen atoms for which the interaction between electron and proton magnetic moments induces a split in two levels of the $1$s ground-state; this kind of atomic clock has been also used by Galileo \cite{Delva:2015kta,Herrmann:2018rva,Delva:2018ilu} and ACES \cite{aces}. The sum of the angular moments of electron and proton spins, ${\bf F}={\bf I}+{\bf s}$, gives the singlet $F=0$ and the triplet $F=1$. The energy difference between the states $|F=1\rangle$ and $|F=0\rangle$, $\Delta E=E_{1}-E_{0}$, is generated by the magnetic moments of the particles: the configuration with parallel magnetic moments of two spins (triplet) corresponds to a higher energetic level than the antiparallel case (singlet). The energy split reads $\Delta E=\frac{4}{3}g_e g_p \alpha^2 \frac{m_e}{m_p}E_0\sim 5.8\times 10^{-6}$eV, where $g_e=2$ and $g_p=5.586$ are the gyromagnetic factors of the electron and proton, respectively, $m_p\simeq 1836 m_e$ is the proton mass with respect to electron mass, and $E_0=-13.6$eV is the hydrogen ground-state energy. The corresponding frequency and wavelength are $\nu_0=1420$MHz and $\lambda_0=21.1$cm, respectively. If the system is prepared in the lower energy state, then the absorption of photons with this wavelength would induce a spin-flip transition increasing the population of hydrogen atoms in the state $|F=1\rangle$.

This hyperfine splitting of the hydrogen atom has been measured to a very high precision \cite{hellwig,essen,Dupays:2003zz}, $\nu_{\rm exp}=(1420\pm9\times 10^{-10})$MHz, with fractional uncertainty  $\Delta\nu_{\rm exp}/\nu_0\simeq 6.34\times 10^{-13}\,.$ Remarkably, the current precision is very suitable for our novel experimental proposal since the fractional error lies outside the Earth observational window in Eq.\eqref{freq-window}, meaning that any detectable effect in this range will be always larger than the experimental error, thus giving a clear signature of new physics; see Fig. \ref{fig12} for an illustration\footnote{Note that we do not discuss how to bring the satellite in circular orbit at $r_B=\frac{3}{2}r_A$ as it goes beyond the scope of this Letter. It is worthwhile mentioning that, typically spacecraft tracking is done by Doppler measurement but using such a method would correspond to using $\delta=0$ to locate the satellite which would ruin the experiment from the beginning. Therefore, other kind of methods should be used, like for instance range measurements, but one has to make sure that the effect induced by modified gravity to put the satellite in orbit is always negligible with respect to the effect one wants to measure with the experimental device on the satellite.}. Note also that a frequency of $1420$MHz falls into the microwave region of the electromagnetic spectrum, implying that this kind of photons can penetrate the atmosphere and reach the satellite.

Although the accuracy reached with hydrogen atoms is already sufficient, more accurate clocks made up of different atoms exist, like $^{87}$Rb and $^{133}$Cs, whose fractional uncertainties are of the order of $10^{-14}$ and $10^{-16},$ respectively \cite{bize}. However, hydrogen atoms are more stable, indeed they are characterized by a smaller fractional stability of the order of $10^{-14}$ at $1$ sec. Since any experiment is always affected by some statistical noise, only after many measurements one can obtain a reliable result, and the stability tells us how many measurements are needed in order to trust the result. Therefore, hydrogen atoms are more stable in the sense that a sufficient level of precision can be obtained more quickly as compared to other atoms emitting in the microwave \cite{optical}.

It is worthwhile mentioning that for future experiments, optical atomic clocks might become even more suitable due to their higher accuracy and stability, indeed in this case the fractional frequency go down to $10^{-18}\,,$ while the fractional stability is of the order of $10^{-17}$ at $1$ sec \cite{optical}. See Table II for a list of possible atomic clocks with corresponding fractional uncertainties.

\begin{table}[t]
	\caption{Several atomic clocks with corresponding fractional frequency and fractional stability at $1$ sec.}
	\centering
	\begin{tabular}{p{0.25\linewidth}p{0.34\linewidth}p{0.24\linewidth}}
		\toprule[1.2pt]\midrule[0.6pt]
		atom   & fractional uncertainty & fractional stability\\
		\midrule[0.6pt]
		$^1$H & $10^{-13}$ & $10^{-14}$ \cite{hellwig,essen,Dupays:2003zz}  \\
		$^{87}$Rb & $10^{-14}$ & $10^{-12}$ \cite{bize}\\
		$^{133}$Cs & $10^{-16}$ & $10^{-13}$ \cite{bize}  \\
		$^{87}$Sr (optical) & $10^{-18}$ & $10^{-17}$ \cite{optical}  \\
		\midrule[0.6pt]\bottomrule[1.2pt]
	\end{tabular}
\end{table}
\section{Applications}

\subsection{Yukawa-like corrections}

We now wish to consider some application in the framework of extended theories of gravity. Note that for any generalization beyond GR, we would expect that the modification to the cross-term in Eq.\eqref{polar-metric}, related to the rotation, is always subdominant as compared to the static contribution. Therefore, for all the considered models we can always neglect the modification to the cross-term induced by new physics.

As a first application, let us consider Yukawa-like corrections to the Newtonian potential:
\begin{equation}
\Phi(r)=-\frac{Gm}{r}\left(1+\beta \,e^{-r/{\lambda}}\right)\,,
\label{yukawa-pot}
\end{equation}
where $\beta$ and $\lambda$ are two free constant parameters.
In screening mechanisms, the extra term is related to an extra propagating scalar field whose profile is governed by the Poisson equation $\nabla^2\phi=\partial V_{\rm eff}/\partial \phi$, where the form of the effective potential depends on the kind of screening model. Following Ref. \cite{sakstein}, for the chameleon screening \cite{veltman,brax} one has $V_{\rm eff}=V+V_{I}$, where $V=\Lambda^4(1+\Lambda^{n}/\phi^n)$, with $\Lambda$ of the order of the cosmological constant $\sim 10^{-12}$GeV so that the field $\phi$ can drive the cosmic acceleration; whereas $V_I=\alpha \rho_m \phi/M_{p}$ describes the interaction of the chameleon field with the matter density $\rho_m$, with $\alpha$ being a coupling constant. In the symmetron model \cite{symmetron}, instead, one has $V=\lambda \phi^4/4!-\mu^2\phi^2/2$ which is the typical Higgs-like quartic self interaction potential, while the coupling with matter is given by $V_I=\alpha\rho_m\phi^2/2M_{p}$. Screening effects lead to the gravitational potential \eqref{yukawa-pot}, with the replacement $\beta\to 2\alpha^2\left(1-\frac{m_s}{m}\right)$, where $m_s\equiv m(r_s)$ is the mass inside the screen radius $r_s$. The Yukawa term turns out to be strongly suppressed as $m_s\approx m$, but becomes enhanced for $r_s \ll R$, where $R$ is the radius of the source with mass $m=m(R)$ \footnote{Very interestingly, the screening mechanisms can violate the weak equivalence principle since one can define a scalar charge $Q_i=m_i\left(1-m_i(r_s^i)/m_i\right)$, where the index $i$ labels different constituents, see Ref.\cite{Hui:2009kc}. Therefore, the force exerted by an external chameleon field is not universal but depends on the structure of the constituents: ${\bf F}_{i-\phi}=\alpha Q_i \nabla \phi$.}.
%
Furthermore, the modified potential (\ref{yukawa-pot}) arises also in other different contexts. For instance, when a Lagrangian like $\mathcal{L}\sim R+\alpha R^2$ is chosen, the corresponding weak-field metric potential has $\beta=1/3$ and $\lambda=\sqrt{3\alpha}$ corresponds to the wave-length of an extra scalar degree of freedom in the gravity sector. A similar behaviour for the gravitational potential also manifests in some models of dark-matter induced effects on the gravity sector \cite{Croon:2017zcu}, but in this case the correction can also acquire an opposite sign and give a repulsive effect. It is worthwhile mentioning that very recently tests of the type of modifications in Eq.\eqref{yukawa-pot} with satellite laser-ranging LARES, LAGEOS, and LAGEOS2 have been discussed in Ref.\cite{ciufolini}.
%

The static frequency shift in Eq.\eqref{static-rot} for the potential \eqref{yukawa-pot} turns out to be non-vanishing when evaluated at $r_{B}=\frac{3}{2}r_A,$ unlike in GR, indeed it reads:
\begin{equation}
\delta_{\rm stat}(r_A)=\frac{Gm\beta}{2r_A}\left(e^{-\frac{r_A}{\lambda}}-e^{-\frac{3}{2}\frac{r_A}{\lambda}}-\frac{r_A}{\lambda}e^{-\frac{3}{2}\frac{r_A}{\lambda}}\right).
\label{yukawa-stat-shift}
\end{equation}
The shift in Eq.\eqref{yukawa-stat-shift} falls down in the Earth observational window \eqref{window-earth} provided $\lambda \gtrsim 1.5\times 10^6\,{\rm m}$, or in other words $\alpha \gtrsim 2.25\times 10^{12}\,{\rm m}^2\,.$  On the other hand, if no effect is detected within the observational window, then we can put the following constraint on the new physical scale: $\alpha\lesssim 2.25\times 10^{12}\,{\rm m}^2$. We can also ask what happens with other gravitational sources. For instance, if we assume the observer $A$ is sitting on the Moon, which might be feasible in the future, the bound turns out to be $\alpha\lesssim 1.0\times 10^{10}{\rm m}^2$. Very interestingly, such a constraint definitely improves the one coming from Gravity Probe B \cite{Everitt:2015qri}, $\alpha \lesssim 5.0\times 10^{11}{\rm m}^2,$ which has been the best realized satellite experiment so far \cite{Naf:2010zy}. It is also worth emphasizing that the best laboratory bound on modification of Newton's law comes from torsion balance experiments performed on Earth, indeed the E\"ot-Wash experiment \cite{Kapner:2006si} provides $\alpha \lesssim 10^{-10}\,{\rm m}^2$ \cite{Naf:2010zy,Chen:2019stu}.

\subsection{Power-like corrections}

We now consider a modification of the Newtonian potential which scales as negative powers of the radial coordinate:
\begin{equation}
\Phi(r)=-\frac{Gm}{r}\left(1+\frac{\Theta^\xi}{r^\xi}\right)\,
\label{non-comm-pot}
\end{equation}
%
where $\Theta$ and $\xi$ are free parameters. This form of the potential appears in Vainshtein \cite{Vainshtein} screening, as well as in models of discrete spacetime \cite{Snyder:1946qz} which are common to several approaches to quantum gravity, for instance some low energy limits of string-theory predict a quantized spacetime structure \cite{Doplicher:1994tu} or Brane-World Gravity \cite{marteens}. As an example, if we consider the context of non-commutative geometry where $\xi=2$ \cite{Chaichian:2007we}, the frequency shift at $r_{B}=3/2r_A$ reads $\delta=(89/88)Gm\Theta^2/r^3$. By comparing with the Earth observational window in Eq.\eqref{window-earth}, we can easily find the following upper bound on the deformation parameter $\Theta\lesssim 2.5\times 10^5{\rm m}\,.$ Also in this case, if we assume the observer $A$ on the Moon we can get better constraints:
$\Theta\lesssim 5.16\times 10^{3}\,{\rm m}\simeq 1.0\times 10^{-12}\,{\rm GeV}^{-1},$
which would improve the current experimental bounds which are of the order of $(10^{-10}-10^{-11})\,{\rm GeV}^{-1}$ \cite{Kanazawa:2019llj}.

\section{Summary and conclusions}\label{conclus-sec}

In this Letter, we have proposed a novel satellite experiment aimed to further test Einstein's GR and better constrain extended gravity models. We have found a peculiar source-satellite configuration for which the static contribution to the frequency shift is vanishing in the case of GR. Therefore, we have determined the observational window in which any kind of detectable effect would imply the existence of new physics beyond either Einstein's GR or special relativity. The chosen experimental device is a system of hydrogen atoms which can be excited through photons with a frequency of $1420$MHz, corresponding to the $21.1$cm line. We have shown that this system can work as a switch for probing  models of modified gravity, in particular the screening mechanism invoked to circumvent Solar system and laboratory tests. Remarkably, in the case of thin-shell solution and exponential potential for the chameleon field, the deviation from Newtonian gravity due to screening effects is of the order $10^{-9}$ \cite{veltman,waterhouse} for the Earth, and it might be tested with the novel space-based experiment proposed in this Letter. Furthermore, we have pointed out that by considering the Moon as the gravitational source one can in principle improve the constraints coming from Gravity Probe B on Yukawa- and power-like modifications.

Let us finally remark that we have worked in a simplified framework in which the photon is sent radially and the satellite is in circular orbit. Further studies and much more work are required to take into account small deviations from circular orbits, spherical symmetry and all kind of disturbances which might alter the experiment \cite{noi}. However, we believe that such a novel experimental proposal might open a new window of opportunity to explore new regimes that so far have not been reached with space-based Solar system experiments, and therefore allow us a deeper understanding of the real nature of the gravitational interaction, as well as to probe new physics beyond standard theories \cite{Colladay:1998fq,Kostelecky:2003fs,AmelinoCamelia:2008qg}.
\begin{acknowledgements}
	The authors are grateful to Guglielmo M. Tino for useful comments.
\end{acknowledgements}

\end{document}